\documentstyle[aps,prl,epsf]{revtex}

\bibstyle{unsrt}
\begin{document}
\draft


\title{Large-order Perturbation Theory for a Non-Hermitian ${\cal
PT}$-symmetric Hamiltonian}

\author{Carl M. Bender$^1$ and Gerald V. Dunne$^2$}
\address{${}^1$Department of Physics, Washington University, St. Louis MO
63130, USA}
\address{${}^2$Department of Physics, University of Connecticut, Storrs CT
06269, USA}

\date{\today}
\maketitle

\begin{abstract}
A precise calculation of the ground-state energy of the complex
${\cal PT}$-symmetric Hamiltonian $H=p^2+\frac{1}{4}x^2+i\,\lambda\, x^3$,
is performed using high-order Rayleigh-Schr\"odinger perturbation theory.
The energy spectrum of this Hamiltonian has recently been shown to be real using
numerical methods. The Rayleigh-Schr\"odinger perturbation series is Borel
summable, and Pad\'e summation provides excellent agreement with the real energy
spectrum. Pad\'e analysis provides strong numerical evidence that the
once-subtracted ground-state energy considered as a function of $\lambda^2$
is a Stieltjes function. The analyticity properties of this Stieltjes function
lead to a dispersion relation that can be used to compute the imaginary part of
the energy for the related real but unstable Hamiltonian
$H=p^2+\frac{1}{4}x^2-\epsilon\, x^3$.
\end{abstract}

\vskip .5cm

It has been conjectured \cite{bessis} that the spectrum of the complex
Hamiltonian
\begin{eqnarray}
H=p^2+\frac{1}{4}x^2+i\,\lambda\, x^3
\label{e1}
\end{eqnarray}
is real and positive. Although there is no rigorous proof of this conjecture,  
it has been argued \cite{stefan} that the reality and positivity of the spectrum
is a consequence of the ${\cal PT}$ symmetry of $H$. (Recall that the parity
operation acts as ${\cal P}:~p\to-p$ and ${\cal P}:~x\to-x$ and that the
antiunitary time reversal operation acts as ${\cal T}:~p\to-p$, ${\cal T}:~x\to
x$, and ${\cal T}:~i\to-i$.) The notion that ${\cal PT}$ symmetry can replace
the much more restrictive condition of Hermiticity has been studied in the
context of quasi-exactly solvable quantum theories \cite{QES}, new kinds of
symmetry breaking in quantum field theory \cite{PARITY,SUP}, and complex
periodic potentials \cite{bands}. There have been many other instances of
non-Hermitian ${\cal PT}$-invariant Hamiltonians in physics. Energies of
solitons in Toda theories with imaginary coupling have been found to be real
\cite{HOLLOW}. Hamiltonians rendered non-Hermitian by an imaginary external
field have been used to study population biology \cite{Nelson+Shnerb} and to
study delocalization transitions, such as vortex flux-line depinning in type-II
superconductors \cite{Hatano+Nelson}.

In this paper we study the large-order behavior of Rayleigh-Schr\"odinger
perturbation theory for the ground-state energy of the complex
${\cal PT}$-symmetric Hamiltonian (\ref{e1}). Note that this Hamiltonian
describes a $0+1$ dimensional $\phi^3$ field theory, and recall that $\phi^3$
theories were the first quantum field theories in which the divergences of
perturbation theory were studied \cite{hurst}. For the Hamiltonian (\ref{e1}) we
find that the perturbation series for the ground-state energy is divergent but
Borel summable. Furthermore, by studying the numerical properties of the Pad\'e
approximants we infer that the (once-subtracted) ground-state energy considered
as a function of $\lambda^2$ is a Stieltjes function. This is a very strong
result because it implies analyticity in the cut-$\lambda^2$ plane and other
properties. [It is surprising that this Stieltjes condition holds for a complex
Hamiltonian such as (\ref{e1}); the proof that the once-subtracted ground-state
energy of the conventional $\lambda x^{2N}$ anharmonic oscillator is a Stieltjes
function of $\lambda$ makes use of Hermiticity.] We then use these analyticity
properties to establish a dispersion relation that yields the precise
large-order behavior of the perturbation series.

Let us consider the conventional Rayleigh-Schr\"odinger perturbation series
about the ground state ($E_0=\frac{1}{2}$) of the harmonic oscillator $H_0=p^2+
\frac{1}{4}x^2$. The perturbed energy has an asymptotic series representation in
powers of $\lambda^2$ because the perturbation $x^3$ is an odd function of $x$:
\begin{eqnarray}
E(\lambda)-\frac{1}{2} \sim \sum_{n=1}^\infty \, b_n \, \lambda^{2n}.
\label{e2}
\end{eqnarray}
[We have chosen the form of $H_0$ so that the perturbative expansion
coefficients $b_n$ in (\ref{e2}) are integers.]

Using recursion formulas, we can easily generate as many terms as desired in
this expansion. The coefficients $b_n$ alternate in sign, and their magnitude
grows rapidly with $n$. The first 20 values are listed in Table 1. We have
computed enough of the coefficients $b_n$ so that we can fit the leading
large-$n$ behavior as
\begin{eqnarray}
b_n \sim (-1)^{n+1} {60^{n+1/2}\over (2\pi)^{3/2}}
\Gamma\left(n+\frac{1}{2}\right)\left[1-{\rm O}\left(\frac{1}{n}\right)\right].
\label{e3}
\end{eqnarray}
Therefore, although divergent, the series in (\ref{e2}) is Borel summable
\cite{hardy,BO}. Observe that if the factor of $i$ were absent from the
Hamiltonian (\ref{e1}), then the perturbation coefficients $b_n$ would not
alternate in sign and the perturbation series would not be Borel summable.

We have performed a Pad\'e analysis \cite{hardy,BO} on the divergent series for
the once-subtracted ground-state energy $[E(\lambda)-\frac{1}{2}]/\lambda^2$.
Using the first 46 perturbation coefficients $b_n$, we find that for all real
positive $\lambda^2$ the diagonal Pad\'e sequence $P^N_N(\lambda^2)$ is monotone
decreasing with increasing $N$, and the off-diagonal Pad\'e sequence
$P^M_{M+1}(\lambda^2)$ is monotone increasing with increasing $M$:
\begin{eqnarray}
P^0_1<P^1_2<P^2_3<\dots<P^M_{M+1}<\dots<P^N_N<\dots<P^2_2<P^1_1<P^0_0.
\label{seq}
\end{eqnarray}
The results for $\lambda=0.125$ are shown in Table 2. If the inequalities in
(\ref{seq}) hold for all $N$ and $M$ and for all real positive $\lambda^2$, then
it is rigorously true that $[E(\lambda)-\frac{1}{2}]/\lambda^2$ is a Stieltjes
function of $\lambda^2$ \cite{BO}. This means that $[E(\lambda)-\frac{1}{2}]/
\lambda^2$ is analytic in the cut-$\lambda^2$ plane, vanishes as $|\lambda^2|\to
\infty$, and is a Herglotz function of $\lambda^2$. [A function $f(z)$ is said
to be Herglotz if ${\rm Im}\,f(z)$ is positive (negative) when $z$ is in the
upper (lower) plane.] The fact that (\ref{seq}) holds for $0\leq M,\,N\leq23$
provide strong numerical evidence that $[E(\lambda)-\frac{1}{2}]/\lambda^2$
is a Stieltjes function. We stress that this is a much stronger result than
merely saying that the divergent series (\ref{e2}) is Borel summable.

Furthermore, in addition to the inequality in (\ref{seq}), the limits of the two
Pad\'e sequences appear to be identical. Therefore, we can extract values for
the Pad\'e summed energy from the two Pad\'e sequences. The best estimate for
the ground-state energy is obtained by averaging the last diagonal and
off-diagonal Pad\'e approximants. (To obtain an estimate of the ground-state
energy from this average we multiply the average by $\lambda^2$ and add
${1\over2}$.) The results are shown in Table 3 for various values of the
coupling $\lambda$. Previous numerical calculations of the ground-state energy
were obtained by direct numerical integration of the Schr\"odinger equation (see
Ref.~\cite{stefan}); this technique gave a typical accuracy of about five
decimal places. The agreement between the method of numerical integration and
the Pad\'e summation is excellent. Moreover, for $\lambda<{1\over10}$ the Pad\'e
technique provides an accuracy of more than ten decimal places. The agreement is
better for smaller values of $\lambda$, as is expected, because of a faster
convergence rate of the Pad\'e sequence.

\begin{table*}[p]
\caption[t1]{The first $20$ perturbation coefficients $b_n$ in the expansion
(\ref{e2}) of the ground-state energy for the complex ${\cal PT}$-symmetric
Hamiltonian (\ref{e1}).}
\begin{tabular}{ll}
$n$  &  $b_n$ \\ \tableline
1 & 11\\
2 & -930\\
3 &  158836\\
4 &  -38501610\\
5 &  11777967516\\
6 &  -4300048271460\\
7 &  1815215203378344\\
8 &  -868277986898581530\\
9 &  464025598165231889260\\
10 & -274145574452876905074540\\
11&  177549419941607942489064216  \\
12&  -125174233315525265299874890500  \\
13&  95490636687662293430130201941400   \\
14&  -78410748996991270671939611723389320   \\
15&  68982408758305101330092396215438198608  \\
16&  -64750700102454900598854145411501140103290    \\
17&  64606224564767863138999679663986778514033420     \\
18&  -68291871149169980983310351232642663615057109020    \\
19&  76244729314392095958565433992857306551429203990968    \\
20&  -89660576791390730762095201994590409692301843683859820
\end{tabular}
\label{table1}
\end{table*}

\begin{table*}[p]
\caption[t2]{The diagonal and off-diagonal Pad\'e sequences $P_N^N(\lambda^2)$  
and $P^N_{N+1}(\lambda^2)$ evaluated at $\lambda=0.125$. Observe the rapid  
convergence and note that the inequalities in (\ref{seq}) are satisfied.}
\begin{tabular}{ccc}
$N$ & $P^N_N$ & $P^N_{N+1}$  \\ \tableline
0  & 11.000000000 & 4.739290085 \\
1  & 7.039037169  &   5.696806799 \\
2  & 6.347866015  &   5.947600655 \\
3  & 6.168265727  &   6.026389220 \\
4  & 6.110857028  &   6.054574069 \\
5  & 6.089906566  &   6.065678176  \\
6  & 6.081499968  &   6.070392205 \\
7  & 6.077873385  &   6.072516805 \\
8  & 6.076216002  &   6.073522627 \\
9  & 6.075421823  &   6.074018882 \\
10 & 6.075025816  &   6.074272525 \\
11 & 6.074821510  &   6.074406195 \\
12 & 6.074712942  &   6.074478558 \\
13 & 6.074653729  &   6.074518675  \\
14 & 6.074620680  &   6.074541394 \\
15 & 6.074601848  &   6.074554510 \\
16 & 6.074590917  &   6.074562214 \\
17 & 6.074584462  &   6.074566813 \\
18 & 6.074580592  &   6.074569597  \\
19 & 6.074578237  &   6.074571306 \\
20 & 6.074576787  &   6.074572368 \\
21 & 6.074575882  &   6.074573036 \\
22 & 6.074575311  &   6.074573460 \\
\end{tabular}
\label{table2}
\end{table*}

\begin{table*}[p]
\caption[t3]{The ground-state energy for the Hamiltonian (\ref{e1}) for various
values of the coupling $\lambda$; the ground-state energy was computed by Pad\'e
summation and by direct numerical integration. The Pad\'e sequences were
computed for the once subtracted energy $[E(\lambda)-\frac{1}{2}]/\lambda^2$.
The diagonal Pad\'e energy refers to the energy extracted from the diagonal
Pad\'e sequence $P_N^N(\lambda^2)$, and the off-diagonal Pad\'e energy refers to
the energy extracted from the off-diagonal Pad\'e sequence $P^N_{N+1}(
\lambda^2)$. The best estimate for Pad\'e energy is the average of the diagonal
and off-diagonal values.}
\begin{tabular}{ccccc}
$\lambda$  & Diagonal Pad\'e energy &Off-diagonal Pad\'e energy&Pad\'e energy  
& Numerical energy \\ \tableline
0.015625    & 0.50263 & 0.50263 & 0.50263 & 0.50263\\
0.03125     & 0.50998 & 0.50998 & 0.50998 & 0.50998\\
0.0625      & 0.53393 & 0.53393 & 0.53393 & 0.53393\\
0.125       & 0.59492 & 0.59492 & 0.59492 & 0.59492\\
0.25        & 0.71305 & 0.71284 & 0.71295 & 0.71294\\
0.5         & 0.91445 & 0.89035 & 0.90240 & 0.90026\\
1.0         & 1.40007 & 1.05817 & 1.22912 & 1.16746\\
2.0         & 3.16075 & 1.14032 & 2.15053 & 1.53078
\end{tabular}
\label{table3}
\end{table*}

The above Pad\'e analysis provides strong evidence that the once-subtracted
ground-state energy is analytic in the cut-$\lambda^2$ plane. Thus, we can
derive a dispersion relation in the expansion parameter $\lambda^2$ to deduce
the leading behavior of the imaginary part of the energy for negative
$\lambda^2$. Physically, this means that we can compute the imaginary part of
the energy (and hence the decay width) of the unstable ground state of the
{\it real} Hamiltonian
\begin{eqnarray}
H=p^2+\frac{1}{4}x^2-\epsilon x^3.
\label{e4}
\end{eqnarray}
Note that the ambiguity in the choice of the sign of the coupling $\epsilon$
corresponds to choosing the sign of $i$ in (\ref{e1}). This has no effect on the
decay width; the sign simply distinguishes the direction (left or right) in
which the potential in (\ref{e4}) is unstable.

In the $t=\lambda^2$ plane there is a cut along the negative $t$ axis, and in  
the standard way \cite{BW,simon,zinn} the $b_n$ coefficients are related to
the discontinuity across the cut by the exact formula
\begin{eqnarray}
b_n=\frac{1}{\pi} \int_0^\infty\, \frac{dt}{t} {D(-t)\over t^n},
\label{e5}
\end{eqnarray}
where $D(-t)$ $(t>0)$ is the imaginary part of $E(\lambda)-\frac{1}{2}$,
evaluated with $\lambda^2$ negative. From the growth estimate (\ref{e3}) we
deduce that
\begin{eqnarray}
D(-t)\sim-{e^{-\frac{1}{60 t}}\over2\sqrt{2\pi\,t}}\left[1+{\rm O}(t)\right]
\qquad(t\to0^+).
\label{e6}
\end{eqnarray}
Thus, the leading contribution (for small $\epsilon$) to the imaginary part of
the energy for the unstable ground state of the Hamiltonian (\ref{e4}) is
\begin{eqnarray}
{\rm Im}[E(\epsilon)]\sim{\exp(-\frac{1}{60\epsilon^2})\over (2\pi)^{3/2}\,
\epsilon}\qquad(\epsilon\to0^+).
\label{e7}
\end{eqnarray}

There are several ways to check this result. First, it agrees with a direct
leading-order WKB calculation \cite{chemists} of the imaginary part of the
energy of the unstable ground state of the real Hamiltonian (\ref{e4}). Second,
applying the ``bounce'' method \cite{coleman} to the real unstable Hamiltonian
(\ref{e4}) we find that
\begin{eqnarray}
{\rm Im}[E(\epsilon)]_{\rm bounce}\sim c\, S_0^{1/2}\,\exp(-S_0)
\qquad(\epsilon\to0^+),
\label{e77}
\end{eqnarray}
where the action $S_0$ of the bounce solution is given by
\begin{eqnarray}
S_0=2\int_0^{\frac{1}{4\epsilon}}dx\, \sqrt{\frac{1}{4}x^2-\epsilon\, x^3}=
\frac{1}{60 \epsilon^2}
\label{e8}
\end{eqnarray}
and $c$ is a constant (whose determination requires the computation of a
fluctuation determinant).

Finally, the answer in (\ref{e7}) is in agreement with the variational
perturbation theory analysis in Ref.~\cite{kleinert}. In fact,
Ref.~\cite{kleinert} contains a higher-order WKB expression for ${\rm Im}[E(
\epsilon)]$. Inserting this higher-order WKB result into the dispersion relation
(\ref{e5}), we obtain a WKB-based prediction for the corrections to the
leading-order growth of the $b_n$ coefficients given in (\ref{e3}):
\begin{eqnarray}
b_n^{\rm WKB} &\sim& (-1)^{n+1}{60^{n+1/2}\over(2\pi)^{3/2}}\Gamma\left(n+
\frac{1}{2}\right)\left[1-{169\over120(n-\frac{1}{2})}-{44507\over28800(n-
\frac{1}{2})(n-\frac{3}{2})}\right.\nonumber\\
&&\qquad\left.-{9563539\over1920000(n-\frac{1}{2})(n-\frac{3}{2})(n-\frac{5}{2}
)}-{189244716209\over 8294400000
(n-\frac{1}{2})(n-\frac{3}{2})(n-\frac{5}{2})(n-\frac{7}{2})}\right.\nonumber\\
&&\qquad -\left.{42943442679817\over 331776000000
(n-\frac{1}{2})(n-\frac{3}{2})(n-\frac{5}{2})(n-\frac{7}{2})(n-\frac{9}{2})}
\right.\nonumber\\
&&\qquad -\left.{342541916236654541\over 398131200000000
(n-\frac{1}{2})(n-\frac{3}{2})(n-\frac{5}{2})(n-\frac{7}{2})(n-\frac{9}{2})
(n-\frac{11}{2})}\right.\nonumber\\
&&\qquad -\left.{933142404651555165943\over 143327232000000000
(n-\frac{1}{2})(n-\frac{3}{2})(n-\frac{5}{2})(n-\frac{7}{2})(n-\frac{9}{2})
(n-\frac{11}{2})(n-\frac{13}{2})}-\dots \right].
\label{e9}
\end{eqnarray}
With these higher-order corrections, this growth estimate of the $b_n$
coefficients is spectacularly accurate. For example,
\begin{eqnarray}
{b_{46}^{\rm WKB}\over b_{46}}=1.00000000807.
\label{e10}
\end{eqnarray}

To conclude we note that the strategy employed here to relate the large-order  
Rayleigh-Schr\"odinger perturbation theory coefficients of a stable (and
Borel-summable) problem to the imaginary part of the energy of an unstable
(and Borel-nonsummable) problem is familiar from the quartic double-well
potential $H=p^2+\frac{1}{4}x^2+g x^4$, which is stable when $g>0$ and
unstable when $g<0$ \cite{langer,BW,zinn}. The novelty in this paper is that  
we begin with a {\it complex} Hamiltonian $H=p^2+\frac{1}{4}x^2+i\lambda x^3$  
which, despite being non-Hermitian, nevertheless appears to be stable in the
sense that it has a real and positive (and discrete) energy spectrum and a
Borel-summable perturbation expansion for the ground-state energy. We can then  
relate the large-order perturbation coefficients to the imaginary part of the  
energy of an unstable state of the real but unstable Hamiltonian
$H=p^2+\frac{1}{4}x^2-\epsilon x^3$. It is interesting to note that the
quartic case is relevant to the physics of instantons \cite{polyakov,coleman}
while the cubic case is relevant to `bounces' in scalar field theories
\cite{coleman} and to string perturbation theory \cite{gross}.

\section* {ACKNOWLEDGEMENT}
\label{s6}

We are grateful to the U.S.~Department of Energy for financial support.


\begin{references}
\bibitem{bessis} D.~Bessis, private discussion. This problem originated from  
discussions between Bessis and J.~Zinn-Justin, who was studying Lee-Yang
singularities using renormalization group methods.
\end{references}
\end{document}